\documentclass[prl, aps, 12pt, showkeys, showpacs,tightenlines] {revtex4}
\usepackage{amsmath}
\usepackage{epsfig}
\usepackage{hyperref}
\begin{document}
\title{Electric Current Fluctuations, Entropy and Ionic Conductivity}
\author{Yong-Jun Zhang}
\email{yong.j.zhang@icloud.com, yong.j.zhang@gmail.com}
\affiliation{Science College, Liaoning Technical University, Fuxin, Liaoning 123000, China}

\begin{abstract}
This paper reports a relation between ionic conductivity and electric current fluctuations. The relation was derived using statistical analysis and entropy approach. The relation can be used to calculate ionic conductivity.
\end{abstract}
\maketitle

\section{introduction}
For non-equilibrium phenomena, many relations have been proposed \cite{reports, review, Alemany} about the current fluctuations. The first relation being proposed is fluctuation theorem \cite{Evans} which was demonstrated experimentally \cite{Wang} and is shown \cite{Searles} to be consistent with Green-Kubo relations \cite{Green, Kubo}. Fluctuations is closely related to probability distribution. The Gaussian distribution has been observed in simulation \cite{Evans} and experiment \cite{Seitaridou}. In fact, the Gaussian distribution can be obtained by statistical analysis if it arises from the Binomial distribution.

When a weak external force is applied, the current fluctuations will be the same except that the mean value will shift by an amount proportional to the external force. One example is ionic conduction, see Fig. \ref{ions}. Here the external force is electric field. For a given electric field, the larger the electric current, the larger the entropy production rate will be. Thus the electric field always tends to increase the electric current in order to maximize the entropy production \cite{Beretta5,Paltridge2,Ziegler,MEPP,Dewar}. This tendency must be counterbalanced. A competition has been identified between the entropy production rate and the current fluctuations \cite{ion_conductivity, applications}. Specifically, the competition is between environment entropy and system entropy which is to describe the intensity of the current fluctuations.


In this paper, we study the ionic conduction with both statistical analysis and entropy competition to produce a fluctuation relation. A time parameter $\Delta t$ is introduced into the relation for the convenience of simulation and experiment.

\section{Statistical analysis and entropy}
\begin{figure}[htbp]
  \begin{center}
    \mbox{\epsfxsize=4.0cm\epsfysize=4.0cm\epsffile{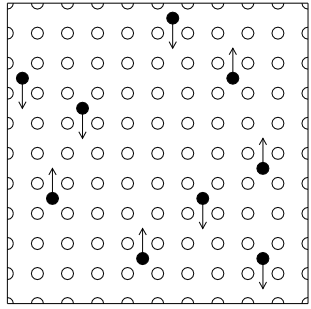}
        }
  \end{center}
\caption{
A simple solid ionic conductor. The open circles are the non-movable lattice ions. The filled circles are the movable interstitial ions that can jump from site to site. Here, only jumping up and jumping down are considered. 
\label{ions}}
\end{figure}

Now let us derive the ionic electrical conductivity. We will do this in a new way different from the ways in work \cite{ion_conductivity, applications}. 
Fig. \ref{ions} shows a solid ionic conductor in the equilibrium state.
The conductor has $N$ interstitial ions, which during a time interval $\Delta t$ make $N_{\Delta t}$ jumps,
\begin{equation}\label{1}
	N=nV,\ \ \ \ \        N_{\Delta t}=nV\frac{\Delta t}{\tau},
\end{equation}
where $n$ is the ion number density, $V$ is the conductor volume and $\tau$ is the mean jump time.
Of all the jumps, $N_{\Delta t}/2+k$ of them are in the up (+) direction and $N_{\Delta t}/2-k$ are in the down (-) direction.
Here $k$ follows a Binomial distribution, which, when $k\ll N_{\Delta t}$, will become a normal distribution,
\begin{equation}\label{Pk} 
        P(k)\propto \exp\left\{-\frac{k^2}{N_{\Delta t}/2}\right\}.
\end{equation}

For a given $k$, there is a corresponding electric current,
\begin{equation} \label{JDeltat}
        J_{\Delta t}=qa\left((N_{\Delta t}/2+k)-(N_{\Delta t}/2-k)\right)=2qak,
\end{equation}
where $q$ is the ion charge and $a$ is the lattice constant. Therefore, Eq. (\ref{Pk}) can be recast as
\begin{equation} 
        P(J_{\Delta t})\propto\exp\left\{-\frac{J_{\Delta t}^2}{2q^2a^2nV\Delta t/\tau}\right\}.
\end{equation}
By introducing a variance,
\begin{equation}\label{lastone} 
	\sigma^2_{J_{\Delta t}}=q^2a^2nV\frac{\Delta t}{\tau},
\end{equation}
it can be written as
\begin{equation} \label{Gaussian}
	P(J_{\Delta t})\propto \exp\left\{-\frac{J_{\Delta t}^2}{2\sigma^2_{J_{\Delta t}}}\right\}.
\end{equation}
If the total number of the microscopic states is $\Omega_{0}$ (for our example, $\Omega_0=2^{N_{\Delta t}}$), the number of microscopic states for a given $J_{\Delta t}$ is
\begin{equation} 
	\Omega(J_{\Delta t})=\Omega_0 P(J_{\Delta t})\propto \Omega_0\exp\left\{-\frac{J_{\Delta t}^2}{2\sigma^2_{J_{\Delta t}}}\right\}.
\end{equation}
 The corresponding entropy is
\begin{equation}\label{S_SJt} 
        S_S(J_{\Delta t})=k_B\ln \Omega(J_{\Delta t})=S_{S0}-\frac{k_BJ_{\Delta t}^2}{2\sigma^2_{J_{\Delta t}}},
\end{equation}
where $S_{S0}$ is a constant. We call this the system entropy \cite{ion_conductivity}. The system entropy is to describe the system in the equilibrium state. The system for this example consists of only those interstitial ions that are responsible for the electric current.

When there exists an external electric field $E$, an environment entropy \cite{ion_conductivity} also has to be introduced, 
\begin{equation} \label{external_E}
        S_E(J_{\Delta t})=S_{E0}+\frac{EJ_{\Delta t}}{T},
\end{equation}
where $T$ is the temperature and $S_{E0}$ is a constant. The environment entropy is introduced by considering $\Delta S=Q/T$ where $Q$ is heat generated by the electric current in the form $Q=EJ_{\Delta t}$. The environment here includes everything except the system. It includes the surroundings and the lattice of the conductor.

Consequently, the overall entropy is
\begin{equation} 
        S(J_{\Delta t})=S_S(J_{\Delta t})+S_E(J_{\Delta t}).
\end{equation}
Therefore the probability distribution is 
\begin{equation}\label{PP} 
        P(J_{\Delta t})\propto\exp\left\{\frac{S(J_{\Delta t})}{k_B}\right\}\propto
        \exp\left\{-\frac{J_{\Delta t}^2}{2\sigma^2_{J_{\Delta t}}}+\frac{EJ_{\Delta t}}{k_BT}\right\}
        \propto\exp\left\{-\frac{1}{2\sigma^2_{J_{\Delta t}}}\left(J_{\Delta t}-\frac{E\sigma^2_{J_{\Delta t}}}{k_BT}\right)^2\right\},
\end{equation}
which indicates that the most probable current is  
\begin{equation} \label{PJt}
	J_{\Delta t}=\frac{E\sigma^2_{J_{\Delta t}}}{k_BT}.
\end{equation}
For a macroscopic system in linear regime ($k\ll N_{\Delta t}$), the most probable current is equal to the average current which is associated with electrical conductivity. Therefore, the electrical conductivity is extracted as
\begin{equation} \label{sigma_e}
	\sigma=\frac{\sigma^2_{J_{\Delta t}}}{k_BTV \Delta t}.
\end{equation}
By Eq. (\ref{lastone}), it becomes $\sigma=\frac{q^2a^2n}{k_BT\tau}$. This is the ionic electrical conductivity that has been obtained in studies \cite{ion_conductivity, applications}. But it has been derived here again in a new way, especially a parameter $\Delta t$ has been introduced.

Although the relation (\ref{sigma_e}) has been derived theoretically from statistical analysis, it can be extended to be suitable for computer simulations. For that, we write
\begin{equation} 
	J_{\Delta t}=\int_0^{\Delta t} J(t) dt,
\end{equation}
which can be observed in simulations. Consequently the electrical conductivity is in the form
\begin{equation}\label{sigma_e2} 
	\sigma=\frac{\langle J_{\Delta t}^2\rangle}{k_BTV\Delta t}.
\end{equation}
This is a novel relation for ionic conduction between electrical conductivity and electric current fluctuations.
Note that $J_{\Delta t}$ and $\langle J_{\Delta t}^2\rangle$ are about the equilibrium state. Also note that there is no requirement for $\Delta t$. But it is safe to choose $\Delta t \gg\tau$. 

\section{conclusion and discussion}
For ionic conductivity, we have presented a relation between electrical conductivity and electric current fluctuations. Electric current fluctuations is described by Gaussian distribution, from which the system entropy is obtained. But when an electric field exists, an additional entropy -- environment entropy -- has to be introduced. The two entropies add. The total entropy gives the current probability distribution. A relation about the electrical conductivity is finally obtained. 
But note that the relation is obtained from the example of ionic conduction which has a unique property: an ion jumps from site to site, which is very different from the way that an ion move through an open space or a liquid. This means that the relation cannot apply directly to ionic conduction in gas or liquid.


{\bf Acknowledgment} 
The author is very grateful to Roberto Trasarti-Battistoni and Bin Zhang for many discussions.

\end{document}